\documentclass[prl,twocolumn,showpacs,preprintnumbers,amsmath,amssymb]{revtex4-1}

\usepackage{graphicx}
\usepackage{dcolumn}
\usepackage{bm}

\begin{document}

\title{Crow instability in trapped Bose-Einstein condensates}
\author{Tapio P. Simula}
\affiliation{School of Physics, Monash University, Victoria 3800, Australia}

\begin{abstract}
We show theoretically that elongated vortex-antivortex dipoles can be created controllably in trapped Bose-Einstein condensates, using known experimental techniques. Vortex dipoles of sufficient length are unstable and cascade into slow vortex rings which ultimately decay via sound emission. This instability of antiparallel vortex line elements, which self-generates Kelvin waves on vortex loops and in trapped atomic gases, may play a role in bridging the Kelvin-wave and Kolmogorov-Richardson cascades of quantum turbulence.
\end{abstract}

\maketitle
Contrails in the sky left behind by aircraft can reveal the pair of counter-circulating wing-tip vortices generated in the wake of the plane \cite{Spreiter1951a}. A large proportion of the energy required to keep an aircraft airborne is consumed in the continuous generation of such wing-tip vortices. These powerful eddies create a bottleneck at airports due to the hazard they impose on aircrafts flying in their vicinity. The Crow instability mechanism seeded by atmospheric turbulence is considered to be a major agent responsible for breaking up these coherent and long-lived wing-tip vortex pairs \cite{Crow1970a,Lewellen2001a}. In water vortex dipoles, an additional short-wave instability has been observed to grow due to a three-dimensional elliptic instability mechanism in an antisymmetric mode together with the long-wave Crow instability \cite{Leweke1998a}. A quantum analogue of the Crow instability \cite{Kuznetsov1995a,Berloff2001a} may occur when bodies are dragged through superfluids creating vortices whose circulation is quantized \cite{Feynman1955a}.

Bose-Einstein condensates (BECs) are versatile quantum liquids exhibiting rich superfluid dynamics \cite{Leggett2006a}. Quantized vorticity and persistent currents are the hallmark of superfluidity in BECs \cite{Matthews1999a,Madison2000a,Abo-Shaeer2001a,Hodby2002a,Ryu2007,Ramanathan2011}. The analog of the Kadomtsev-Petviashvili \cite{Kadomtsev1970a} or `snake' instability of solitons has been observed to lead to generation of vortices and vortex rings in BECs \cite{Anderson2001a,Dutton2001a,Ginsberg2005a,Ruostekoski2005a,Shomroni2009a,Ma2010a}. In the limit of antiparallel vortices it further transforms into the Crow instability \cite{Kuznetsov1995a}. In pancake condensates vortex-antivortex pairs or vortex dipoles can be nucleated via active stimulation \cite{Neely2010a}. For suitable parameter regimes such vortex shedding from a moving obstacle \cite{Inouye2001a} is theoretically predicted to exhibit a B\'enard-von K\'arm\'an vortex street structure \cite{Sasaki2010a}. Vortex dipoles can also form spontaneously via the Berezinskii-Kosterlitz-Thouless mechanism in finite-temperature systems in which entropy may cover the energy cost of pair creation \cite{Simula2006a,Hadzibabic2006a,Clade2009a,Tung2010a,Hung2011a}. Furthermore, a quench through a BEC phase transition facilitates stochastic formation of vortex dipoles via the Kibble-Zurek scenario as observed in recent experiments \cite{Weiler2008a,Freilich2010a}. 

Solitary vortex dipoles have been created and observed experimentally in oblate condensates. However, the three-dimensional Crow instability is manifestly absent in sufficiently flat condensates where axial degrees of freedom play no role in vortex dynamics. Since spontaneous formation of long three-dimensional vortices is suppressed, an active pair-creation method is needed to isolate and observe three-dimensional instabilities of interacting vortex lines. Here we propose such a method to generate long vortex dipoles in trapped quantum degenerate gases and study their superfluid decay dynamics.

\begin{figure*}
\includegraphics[width=1.3\columnwidth]{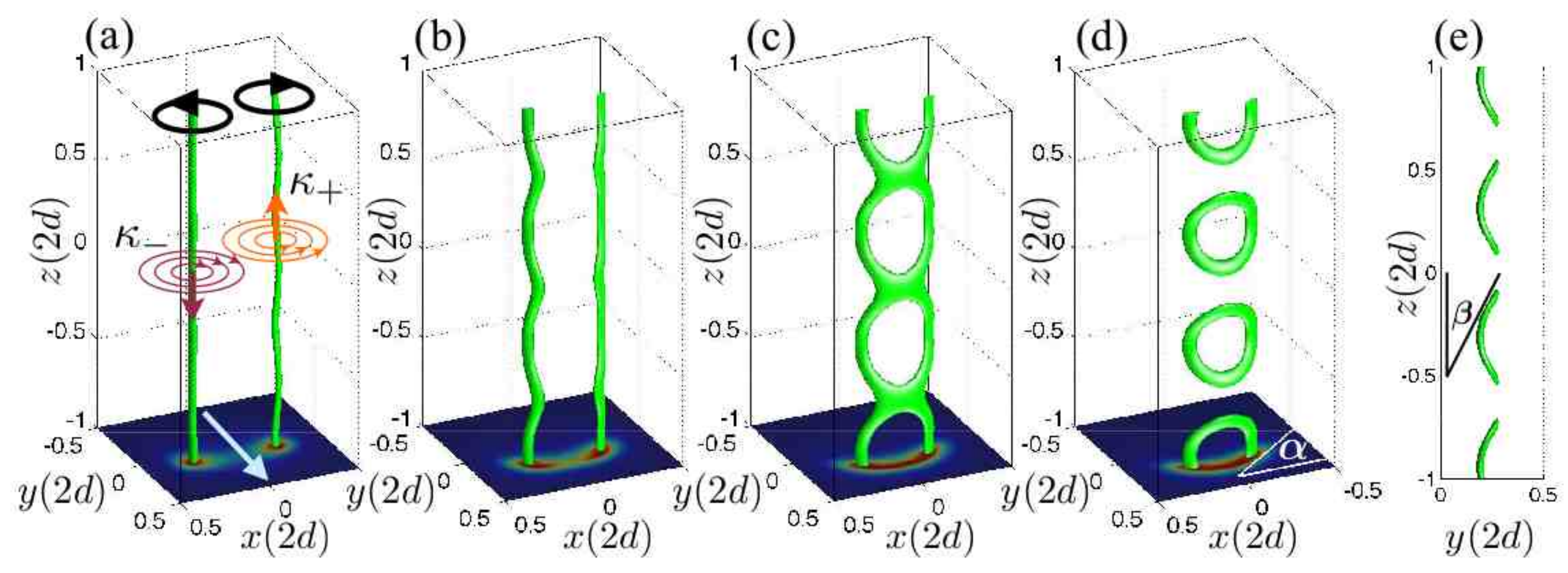}
\caption{Evolution of a sinusoidally perturbed vortex-antivortex dipole. (a) The superflow around the vortices is denoted by the circulation vectors $\kappa_\pm$ and the oriented circles on top of the subfigure indicate the rotational motion of vortex line elements in the direction opposite to the circulating superflow. The pair translates in the $y$ direction shown by the arrow, due to the mutual induction field.  (b) The amplitude of the instability grows. (c) Vortex reconnections occur at locations where vortices come in contact with each other.  (d) The initial vortex pair has broken into a sequence of vortex loops. (e) Side view of frame d. The projection of the vortex lines on the $z=-2d$ plane are shown on the bottom of each subfigure a-d which are plotted, respectively, for times $t/t_0=$0.02, 0.04, 0.05, and 0.06, where the unit of time $t_0=4md^2/\hbar$. Parameters described in the text are $\epsilon = d / 200$ and $k= 5 / d$.}
\label{fig1}
\end{figure*}

Bose-Einstein condensates may be represented by a macroscopic wavefunction $\phi({\bf r},t)$ whose evolution is governed by the Gross-Pitaevskii equation 
\begin{equation}
i\hbar\partial_t \phi({\bf r},t)  = \left(-\frac{\hbar^2\nabla^2}{2m}+V_{\rm ext}({\bf r})+\frac{4\pi \hbar^2 a}{m}|\phi({\bf r})|^2 \right)\phi({\bf r},t), 
\label{GP}
\end{equation}
where the constant $a$ is the $s$-wave scattering length of particles with mass $m$, and $V_{\rm ext}({\bf r})$ is an external potential used to confine and manipulate the atoms. We first consider an initial state wave function $\phi({\bf r},0) = (x+d/2 +\epsilon\cos(kz) +iy)(x-d/2 -\epsilon\cos(kz) -iy)$ which uses Cartesian coordinates to represent two counter-circulating vortex lines separated by distance $d$. The vortices are sinusoidally perturbed with an amplitude $\epsilon$ and a wave vector $k$ in a symmetric mode about their equilibrium positions, see Fig.~1(a).  This vortex dipole generates a mutual induction field which causes the pair to travel with a speed inversely proportional to $d$ in the positive $y$ direction. In addition to the translational motion, each perturbed vortex spins about its own axis. This self-induced spinning motion is due to the curvature of the vortex and comes in the form of Kelvin waves of growing amplitude \cite{Thomson1880a,Simula2008a}. In the Crow instability mode the vortices become phase-locked at a certain angle when the total induced velocity field stops the vortices from spinning. Eventually parts of the vortices overlap, forming a chain of vortex reconnections, shown in Fig.~1(c), which break the vortex dipole into a sequence of vortex loops, see Fig.~1(d). The dynamics illustrated in Fig.1 is obtained by considering a homogeneous ($V_{\rm ext}=0$) non-interacting ($a=0$) system, formally integrating Eq.~(\ref{GP}) and truncating the power series expansion of the resulting matrix exponential to second order, yielding an analytically soluble model. Vortices are visualized in Fig.~1 by plotting isosurfaces of the function $|\phi({\bf r},t)|^2$.

Next we model the full nonlinear dynamics of an elongated vortex dipole embedded in an inhomogeneous background by numerically solving the full Gross-Pitaevskii equation. We consider $N=\int|\phi({\bf r})|^2 d{\textbf r}=8\times10^5$ Bose-Einstein condensed $^{87}$Rb atoms in an anisotropic harmonic potential $V_{\rm ext}({\bf r})=m\omega^2 (x^2+y^2+\lambda_z^2z^2)/2$, where $\lambda_z=\omega_z/\omega = 0.2$, and $\omega = 2\pi\times 100$Hz. The condensate healing length $\xi_0=0.15\mu m$ characterizes the size of the vortex core while for our parameters the harmonic oscillator length $a_0=\sqrt{\hbar/m\omega}= 6.6\xi_0$. We prepare the initial state by phase imprinting two straight $\epsilon = 0$ vortex lines of opposite circulation in the condensate wavefunction. The cores of the vortices are placed at locations $(x_0=\pm d/2,y_0=-2 a_0)$ and are oriented along the $z$ axis. The ends of the dipole are deliberately short-circuited at $z_0=\pm20 a_0$, topologically forming a highly anisotropic vortex ring, shown in Fig.~2(a). Note that vortices described by a single analytical complex function cannot have free ends and will always form closed loops, although in practice quantum and/or thermal fluctuations may provide an effective boundary for the condensate where vortices may terminate. 

The mother vortex ring is propelled forward in the $y$ direction by the self-induced superflow. The inhomogeneous condensate density causes the vortex line elements in the lower density regions to travel faster than those in higher density regions. Due to this differential velocity the ends of the dipole bend, self-generating Kelvin waves which then propagate along the vortex lines symmetrically from both ends toward the center of the condensate, see Fig.~2(a) and Fig.~2(b). These Kelvin waves catalyse the growth of instabilities on the vortex dipole. A train of daughter vortex loops is created when the unstable modes have grown in amplitude to close the gap separating the vortices, igniting reconnections at the nodes of the most unstable modes, shown in Fig.~2(b) and Fig.~2(c). This violent process leaves the daughter vortex loops ringing (carrying Kelvin wave excitations) transforming them into slow vortex rings \cite{Barenghi2006a,Hershberger2010a}. These daughter vortex loops subsequently undergo several recombinations and reconnections and ultimately convert their energy into sound waves. If the size of the generated vortex loops is large enough, they will penetrate the condensate surface. This can result in rotation of the axis of vorticity by 90 degrees, from initially being oriented along the $z$ axis to lying along the $x$ axis, cf. Fig.~2(a) and Fig.~2(d). Table I summarizes the dependence of the instability on the separation $d(\xi_0)$ between the vortices. The wavelength of the fastest growing mode $\lambda_{\tau}(\xi_0)$ and the angles $\beta_\tau$ and $\alpha_\tau$ (cf. Fig.~1(d) and Fig.~1(e)), which in our case are approximately equal, are measured at the time $\tau$ when the vortex dipole first crosses the line $y=z=0$.

\begin{table}
\caption{Instability parameters for different initial vortex dipole separations $d$. Mean distance $d_\tau$ separating the vortex lines at time $\tau$ is shown in parenthesis.}
\begin{ruledtabular}
\begin{tabular}{ccccc}
$d(\xi_0)$ & $\lambda_{\tau}(\xi_0)$ & $\beta_\tau(\angle)$ & $\tau ($ms$)$\\
\hline
26(11) &  20  &  55$^\circ$ & 12\\
20(10) & 14  &  43$^\circ$  & 9\\
12(7) &  12  &  40$^\circ$  & 6\\
9(6)   &  12  &  22$^\circ$  & 4\\
5(4)   &  12  &  18$^\circ$ & 2\\
\end{tabular}
\end{ruledtabular}
\end{table}

\begin{figure*}
\includegraphics[width=1.3\columnwidth]{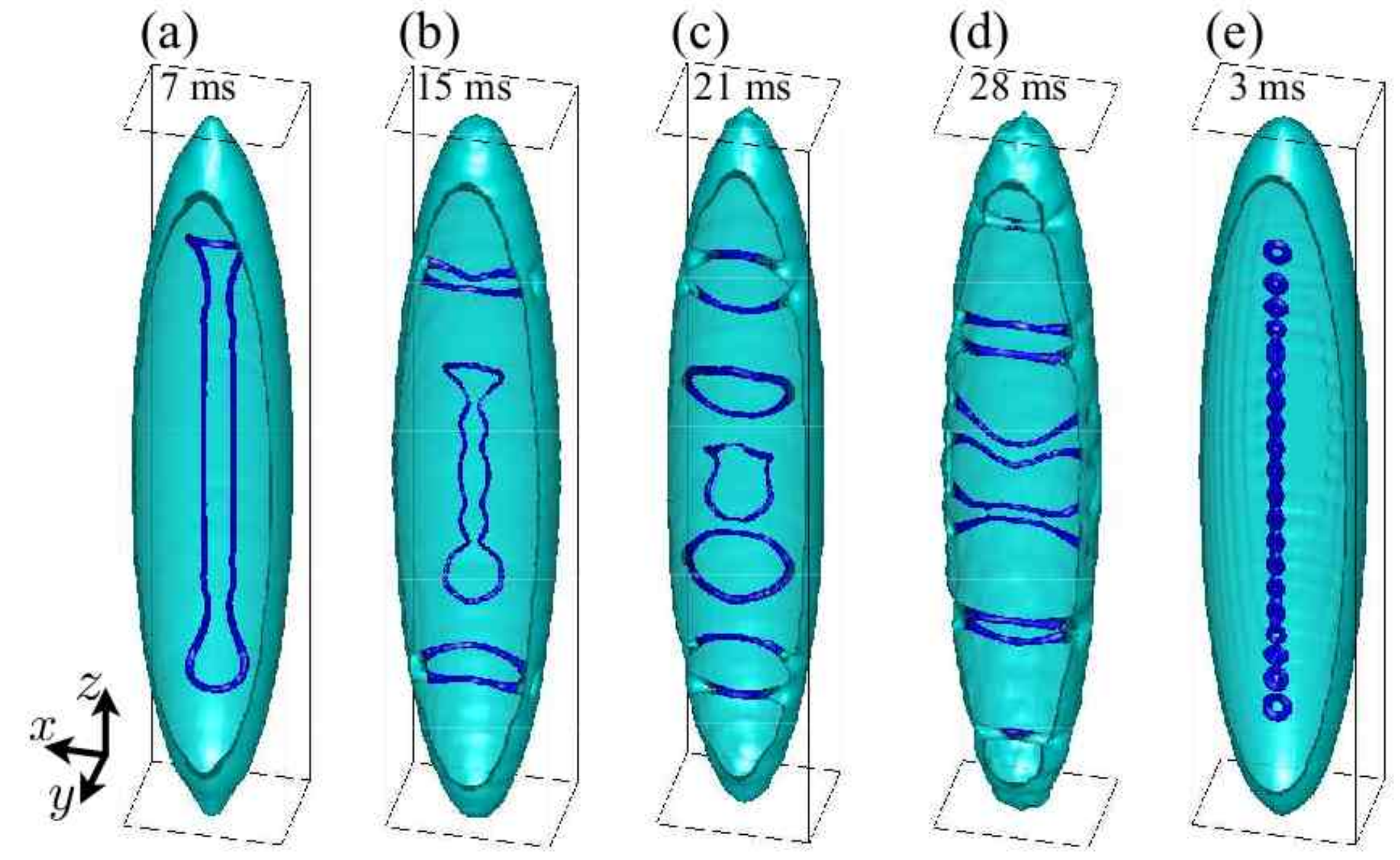}
\caption{Crow instability of an elongated vortex dipole in a harmonically trapped Bose-Einstein condensate for $d=28\xi_0$. (a) Kelvin waves are being generated at the ends of the dipole. (b) Two vortex loops have been pinched off, one from each end of the dipole. (c) Five daughter vortex loops have been produced through reconnection events. (d) The five vortex loops have expanded in diameter appearing as ten vortex lines. The vorticity axis is now aligned along the $x$ axis. (e) A snapshot from a simulation using $d=5\xi_0$. The $x\times y\times z$ dimensions of each rectangular box are $10a_0\times10a_0\times60 a_0$ and the times of the snapshots are marked in the frames.}
\label{fig2}
\end{figure*}

Finally, we propose an experiment to probe the physics described above. We nucleate vortex-antivortex dipoles in the wake of a moving repulsive laser beam which is tuned off-resonance from a suitable atomic transition. This creates a repulsive potential   
\begin{equation}
V ({\bf r},t)= V_0(t) \frac{\sigma_0^2}{\sigma(z)^2}e^{-2r(t)^2/\sigma(z)^2}
\end{equation}
for the condensate atoms. Here we use $r^2=x^2+(y-y_0+vt)^2$, $\sigma(z)=\sigma_0\sqrt{1+(z/z_R)^2}$ and the beam is focused to $\sigma_0=z_R/10$ with Rayleigh range $z_R=22$ $\mu$m and $V_0 = 12.5[1+\tanh(t-\tau)]\hbar\omega$. This laser spoon travels in the condensate for $\tau=6$ ms at a speed $v= 0.68$ mm/s corresponding to a Mach number 0.4 and is then smoothly withdrawn by ramping down the laser power. In Fig.~3a the laser beam is highlighted in the condensate density isosurface and its motion is along the $y$ axis. After turning off the laser beam a vortex dipole is revealed, which nucleated in the wake of the laser, see Fig.~3(b). This vortex dipole becomes spontaneously short circuited at its ends due to the combination of the inhomogeneous condensate density and the curvature in the laser intensity profile. The mother vortex loop breaks into five primary daughter loops, shown in Fig.~3(c) and Fig.~3(d), which undergo multiple reconnections before being converted into sound waves, which become visible as ripples on the condensate surface as time progresses, see frames (e)-(g) in Fig.~3. Increasing the Mach number of the traveling laser spoon increases the number of mother vortex rings produced in the wake of the laser. The large number of condensate and laser parameters readily accessible in experiments enables controlled production of elongated vortex dipoles and daughter vortex loops. 

\begin{figure*}
\includegraphics[width=1.3\columnwidth]{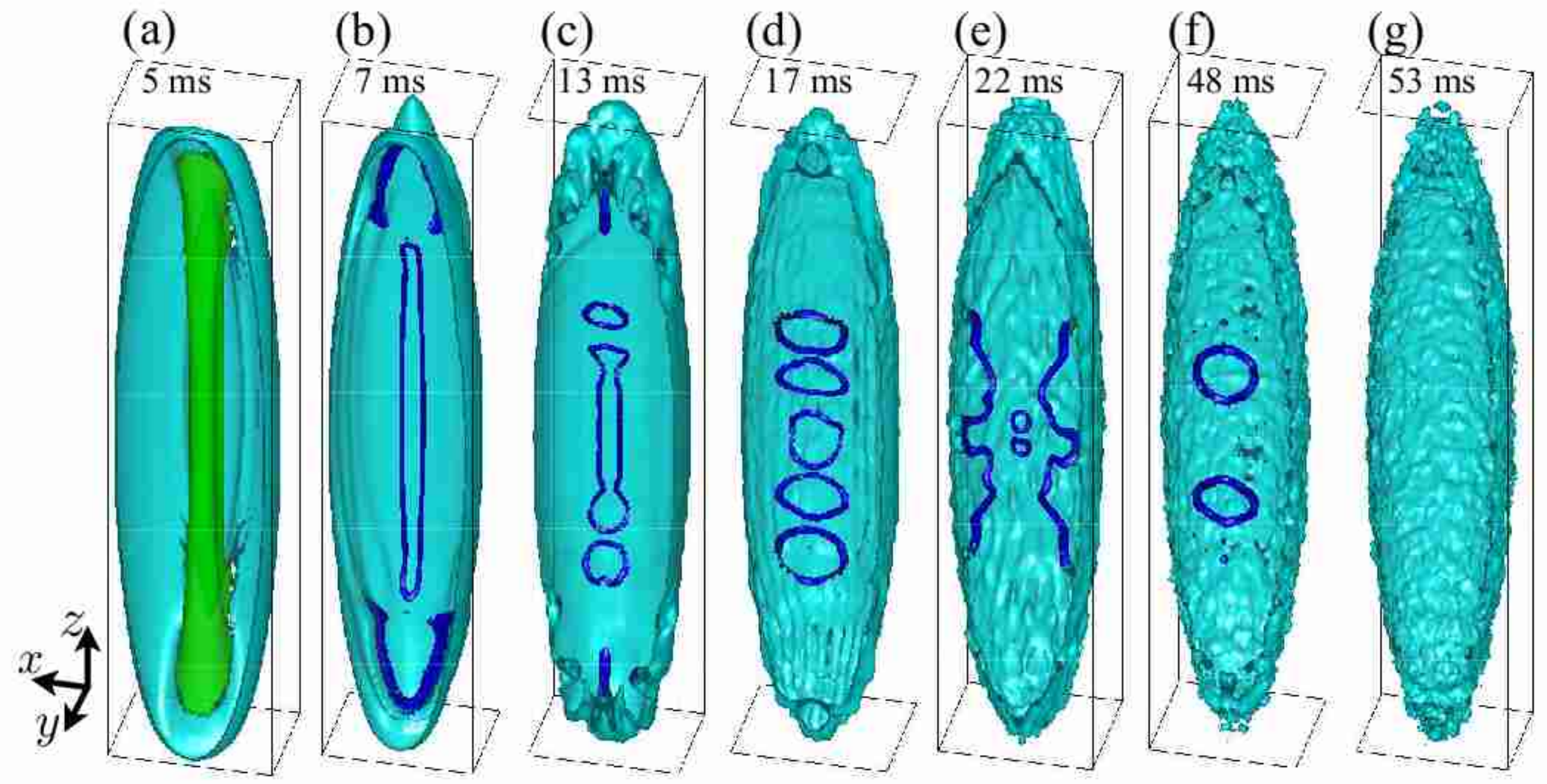}
\caption{Vortex dipole creation and instability in a harmonically trapped Bose-Einstein condensate. (a) A laser beam shown in green translates in the $y$ direction. (b) Vortex dipole has formed in the wake of the laser spoon. (c) Decay of the dipole has begun. (d) Five daughter loops have been created. (e) Daughter vortex loops have been relinked in further reconnections. (f) A snapshot with two vortex rings. (g) Initial vortex dipole has disintegrated creating sound waves seen as ripples on the condensate surface. The $x\times y\times z$ dimensions of each rectangular box are $10a_0\times10a_0\times60 a_0$ and the times of the snapshots are marked in the frames.}
\label{fig3}
\end{figure*}

Vortex dipoles and their dynamics in Bose-Einstein condensates have recently attracted considerable interest \cite{Weiler2008a,Neely2010a,Freilich2010a,Kuopanportti2011a,Rooney2011a}. Experiments have considered effectively two-dimensional systems where the vortex dipoles are long-lived structures. We have shown that in elongated harmonically trapped condensates where axial vortex degrees of freedom are active, vortex dipoles become susceptible to the Crow instability mechanism. Self-generated Kelvin-wave excitations cause vortex dipoles to disintegrate forming a sequence of vortex loops which eventually decay into sound waves. We have proposed and simulated an experiment to study this phenomenon. Due to the high degree of controllability in this vortex-dipole creation method, it could also be used to reproducibly generate loopy vortex states and to study their decay. 

Our results have further implications for decay mechanisms of quantum turbulence in harmonically trapped Bose-Einstein condensates. In uniform systems, a reconnection-dominated Kolmogorov-Richardson cascade has been suggested to be transformed to a Kelvin-wave cascade when vortex-vortex collisions become too infrequent to yield a sufficient vortex reconnection rate. However, the self-generated vortex loop instabilities may be able to sustain vortex reconnections down to the dissipation scale. It will be particularly interesting to apply vortex-dipole nucleation methods to spinor BECs where the evolution of non-Abelian vortex dipoles with fractional charge may lead to drastically different vortex dynamics.

\begin{acknowledgements}
I thank David Paganin for many useful discussions and comments on this manuscript.
\end{acknowledgements}

\end{document}